\documentclass{ws-p8-50x6-00}
\usepackage{feynmf}
\usepackage{amssymb}   

\begin{document}

\title{Jet Physics at HERA}

\author{Hans-Christian Schultz-Coulon}
\address{Universit\"at Dortmund, 44221 Dortmund, Germany\footnote{This 
work was supported by the BMBF, contract no.\ 7D055P.}}

\maketitle

  \abstracts{Jet production in electron-proton collisions at HERA 
  provides a unique testing ground for Quantum Chromodynamics~(QCD).  
  Apart from the determination of the strong coupling constant 
  $\alpha_s$, $ep$~jet data may especially be used to gain insight 
  into the dynamics of the exchanged parton cascade, whose structure 
  is probed by the high-$E_T$ dijet system; thus information on the 
  parton content of the proton and \mbox{(quasi-)}\-real and virtual 
  photons is obtained.  This report touches some of these aspects 
  revealed in recent jet data from the HERA experiments which are
  testing perturbative QCD at the limits of applicability.}

\section{Introduction}

Multi-jet production in deep inelastic $ep$-scattering~(DIS) provides 
special sensitivity to the mechanisms of the strong interaction and can 
hence be used to test the predictions of perturbative Quantum 
Chromodynamics (pQCD) in a rather inimitable way.  This is especially 
true since the HERA experiments deliver data over a large range of 
both the four-momentum transfer, $Q^2$, of the exchanged photon and the 
transverse energy of the observed jets,~$E_T$.

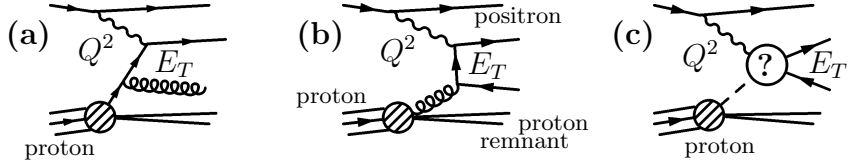
\begin{figure}[t]
\makebox[0.31\textwidth][c]{
  \begin{fmffile}{fmfqcdc}
  \begin{fmfchar*}(75,45)
  \fmfset{wiggly_len}{2.5mm}
  \fmfset{curly_len}{1.5mm}
  \fmfset{arrow_len}{2.0mm}
  \fmfleft{ip,il} 
  \fmfright{oq1,oq2,d1,oq3,xx,d2,yy,oq4,d3,d4,ol}
  \fmf{fermion,tension=5}{ip,vp}
  \fmf{plain}{vp,oq1}  
  \fmf{plain}{vp,oq2}  
  \fmf{photon,label={\large $Q^2$},label.side=right,tension=1,label.dist=1.1mm}{vl,vq}  
  \fmf{fermion,tension=4}{il,vl}
  \fmf{fermion}{vl,ol}
  \fmf{fermion,tension=1}{vp,vg,vq,oq4}
  \fmfblob{.15w}{vp}
  \fmffreeze
  \fmf{phantom}{vg,z1,z2,z3,z4,oq3}
  \fmf{fermion}{vg,vq}
  \fmf{fermion}{vq,oq4}
  \fmffreeze
  \fmf{phantom,label={\large $E_T$},label.side=left,label.dist=1.1mm}{vg,oq3}
  \fmf{gluon}{vg,z4}
  \fmfi{plain}{vpath (__ip,__vp) shifted (thick*(0,2))}
  \fmfi{plain}{vpath (__ip,__vp) shifted (thick*(1,-2))}
  \fmfiv{label={\small proton},label.angle=-114,label.dist=11mm}{c}  
  \fmfiv{label={\large\bf (a)},label.angle=160,label.dist=11mm}{c}  
  \end{fmfchar*}
  \end{fmffile}}\hfill
\makebox[0.35\textwidth][c]{
  \begin{fmffile}{fmfbgf}
  \begin{fmfchar*}(75,45)
  \fmfpen{thin}
  \fmfset{wiggly_len}{2.5mm}
  \fmfset{curly_len}{1.5mm}
  \fmfset{arrow_len}{2.5mm}
  \fmfleft{ip,il} 
  \fmfright{oq1,oq2,d1,oq3,xx,d2,yy,oq4,d3,d4,ol}
  \fmf{fermion,tension=5}{ip,vp}
  \fmf{plain}{vp,oq1}  
  \fmf{plain}{vp,oq2}  
  \fmf{photon,label={\large $Q^2$},label.side=right,tension=1,label.dist=1.1mm}{vl,vq}  
  \fmf{fermion,tension=4}{il,vl}
  \fmf{fermion}{vl,ol}
  \fmf{fermion}{oq3,vg,vq,oq4}
  \fmf{gluon,tension=1}{vp,vg}
  \fmfblob{.15w}{vp}
  \fmffreeze
  \fmf{phantom,label={\large $E_T$},label.side=left,label.dist=1.1mm}{vg,oq3}
  \fmfi{plain}{vpath (__ip,__vp) shifted (thick*(0,2))}
  \fmfi{plain}{vpath (__ip,__vp) shifted (thick*(1,-2))}
  \fmfiv{label={\small proton},label.angle=-156,label.dist=8mm}{c}  
  \fmfiv{label={\small positron},label.angle=35,label.dist=8.5mm}{c}  
  \fmfiv{label={\small proton},label.angle=-28,label.dist=14.5mm}{c}  
  \fmfiv{label={\small remnant},label.angle=-50,label.dist=12mm}{c}  
  \fmfiv{label={\large\bf (b)},label.angle=160,label.dist=11mm}{c}  
  \end{fmfchar*}
  \end{fmffile}}
\makebox[0.33\textwidth][c]{
  \begin{fmffile}{fmfgen}
  \begin{fmfchar*}(75,45)
  \fmfpen{thin}
  \fmfset{wiggly_len}{2.5mm}
  \fmfset{curly_len}{1.5mm}
  \fmfset{arrow_len}{2.5mm}
  \fmfleft{ip,il} 
  \fmfright{oq1,oq2,d1,oq3,xx,d2,yy,oq4,d3,d4,ol}
  \fmf{fermion,tension=5}{ip,vp}
  \fmf{plain}{vp,oq1}  
  \fmf{plain}{vp,oq2}  
  \fmf{photon,label={\large $Q^2$},label.side=right,tension=1,label.dist=1.1mm}{vl,vertex}  
  \fmf{fermion,tension=4}{il,vl}
  \fmf{fermion}{vl,ol}
  \fmf{fermion}{oq3,vertex}
  \fmf{fermion}{vertex,oq4}
  \fmfiv{label={\large $E_T$},label.angle=0,label.dist=10mm}{c}  
  \fmf{dashes,tension=1}{vp,vertex}
  \fmfv{decor.shape=circle,decor.size=.20w,decor.filled=empty,label.dist=-1.2mm,label={\large\bf ?}}{vertex}
  \fmfblob{.15w}{vp}
  \fmffreeze
  \fmfi{plain}{vpath (__ip,__vp) shifted (thick*(0,2))}
  \fmfi{plain}{vpath (__ip,__vp) shifted (thick*(1,-2))}
  \fmfiv{label={\small proton},label.angle=-100,label.dist=10mm}{c}  
  \fmfiv{label={\large\bf (c)},label.angle=160,label.dist=11mm}{c}  
  \end{fmfchar*}
  \end{fmffile}}
  \vspace{-.2cm} 
\caption{Schematic Feynman diagrams for the dijet 
  production mechanism with (a) the QCD-Compton and (b) the 
  boson-gluon fusion process; (c) generic picture for dijet production 
  via photon-parton fusion where the actual interaction mechanism is 
  unknown and to be resolved by the dijet system.  The two scales 
  involved in these processes are indicated as the four-momentum 
  transfer $Q^2$ and the transverse jet energy $E_T$.}
\label{fig=qcdc_bgf}
\vspace{-.6cm}
\end{figure}

In lowest order, multi-jet production in DIS is described by the 
QCD-Compton and the boson-gluon fusion processes
(Figure~\ref{fig=qcdc_bgf}), with the momentum distributions of the 
incoming partons taking part in the hard interaction given by the 
parton density functions (PDFs) of the proton; the $Q^{2}$-evolution 
of the latter is described by the DGLAP equations\cite{DGLAP} which in 
lowest order are equivalent to the assumption of exchanging a strongly 
$k_{t}$-ordered parton cascade, $k_{t}$ being the transverse momentum 
of the partons within the cascade.  However, as illustrated by the 
generic diagram in Figure~\ref{fig=qcdc_bgf}c, to calculate $ep$ jet cross sections 
requires additional terms, such that in certain parts of phase 
space, where $k_{t}$-ordered parton emission is no longer manifest, 
the standard DGLAP approach has to be extended e.g.\ with the concept 
of photon structure, which preserves the perturbative ansatz of the 
DGLAP evolution scheme but results in an artificial violation of the 
$k_{t}$-ordering.  Within this concept, high $E_T$ jets can be 
produced not only by direct processes in which the virtual photon 
interacts as a point-like particle with a parton out of the proton, 
but also by resolved processes where the photon interacts 
hadronically.  In the photoproduction limit ($Q^{2}\rightarrow 0$), 
the cross section is thus sensitive to the parton distributions of 
both the photon and the proton at a scale set by the transverse 
energy of the jets, $E_T$.

The two scales involved in $ep$ jet production, $\sqrt{Q^2}$ and $E_T$, can now 
be used to resolve the structure of the photon-parton interaction in 
general.  Several kinematic regions can be distinguished, depending on 
the absolute and relative sizes of these two scales.  In the regime 
where $Q^2$ and $E_T^2$ are both large, perturbative methods are 
clearly justified and the DGLAP ansatz can be used to extract 
information on the proton structure and the strong coupling 
$\alpha_s$.  Typically, leading-order (LO) Monte Carlo models 
approximate higher order pQCD contributions in this regime by parton 
showers.  If $Q^2\ll E_T^2$ the jet cross sections become sensitive 
to the hadronic structure of the photon.  The region where $Q^2\approx 
E_T^2$ is of special interest, since in this regime the DGLAP ansatz 
is expected to break down for small values of the variable $x_{B}$ 
(Bjorken-$x$) revealing effects of non-$k_{t}$-ordered parton cascades.  
Characteristics expected from other evolution schemes like 
BFKL\cite{BFKL} or CCFM\cite{CCFM} could be observable in this phase 
space regime.  

Any deviations from next-to-leading order (NLO) predictions may, 
however, simply indicate that fixed order calculations beyond NLO are 
needed.  These may also help to resolve other ambiguities associated 
with high $E_{T}$ jet production in DIS: where $Q^{2}$ and $E_{T}^{2}$ 
both are substantially large ($\gg \Lambda_{QCD}^{2}$) there is 
ambiguity in the choice of the renormalization 
scale\footnote{Analogously such ambiguity exists also for the 
factorization scale.  However, since the effects due to the choice of 
the renormalization scale are generally larger, this ambiguity is not 
explicitly mentioned in the context of this paper.}.  It is important 
to know where these ambiguities have a significant effect on the cross 
sections.

\section{Testing pQCD at Large Scales}

\begin{figure}[b]
\vspace{-.2cm}    
\begin{minipage}{0.50\textwidth}
  \flushright{\epsfig{figure=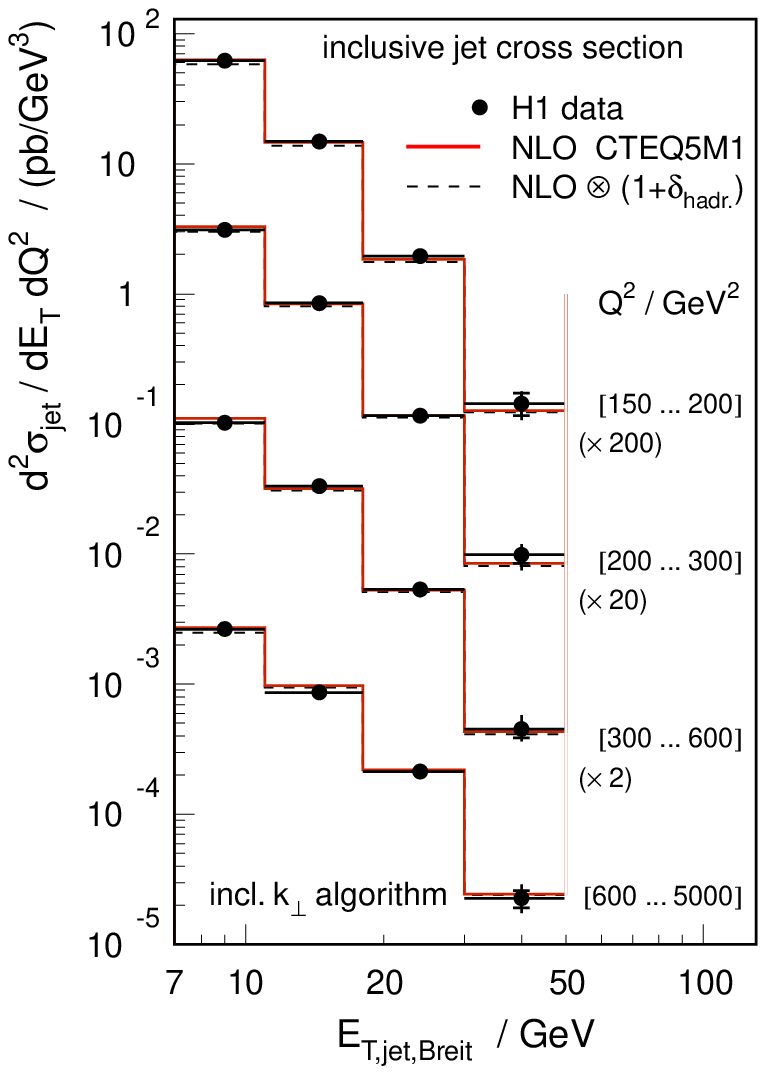,width=\textwidth}}
\end{minipage}\hfill
\raisebox{0.4cm}{
\begin{minipage}{0.48\textwidth}
  \centerline{\epsfig{figure=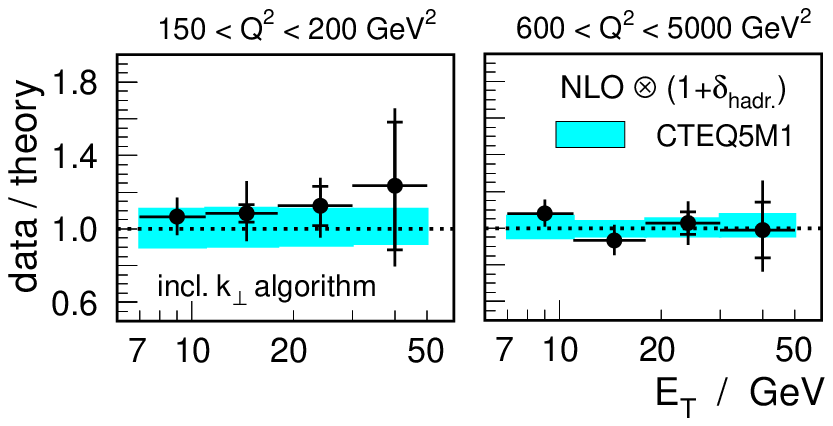,width=\textwidth}}
    \vspace{-.3cm} \parbox{\textwidth}{ \caption{The left figure 
    shows the inclusive jet cross section as a function of the 
    transverse jet energy $E_T$ in different regions of $Q^2$.  Jets 
    are reconstructed in the Breit frame using the inclusive 
    $k_{\perp}$ algorithm.  The data are compared to the NLO pQCD 
    prediction with and without hadronization corrections.  For two 
    of the $Q^{2}$-bins the right figure shows the corresponding 
    ratio of data to the theoretical prediction; the uncertainties 
    due to hadronization effects and the choice of renormalization and 
    factorization scales are indicated by the band.}
  \label{fig=incl_jets_kt}}
\end{minipage}}
\vspace{-.8cm}
\end{figure}

\begin{figure}[t]
\begin{minipage}{0.55\textwidth}
  \centerline{\epsfig{figure=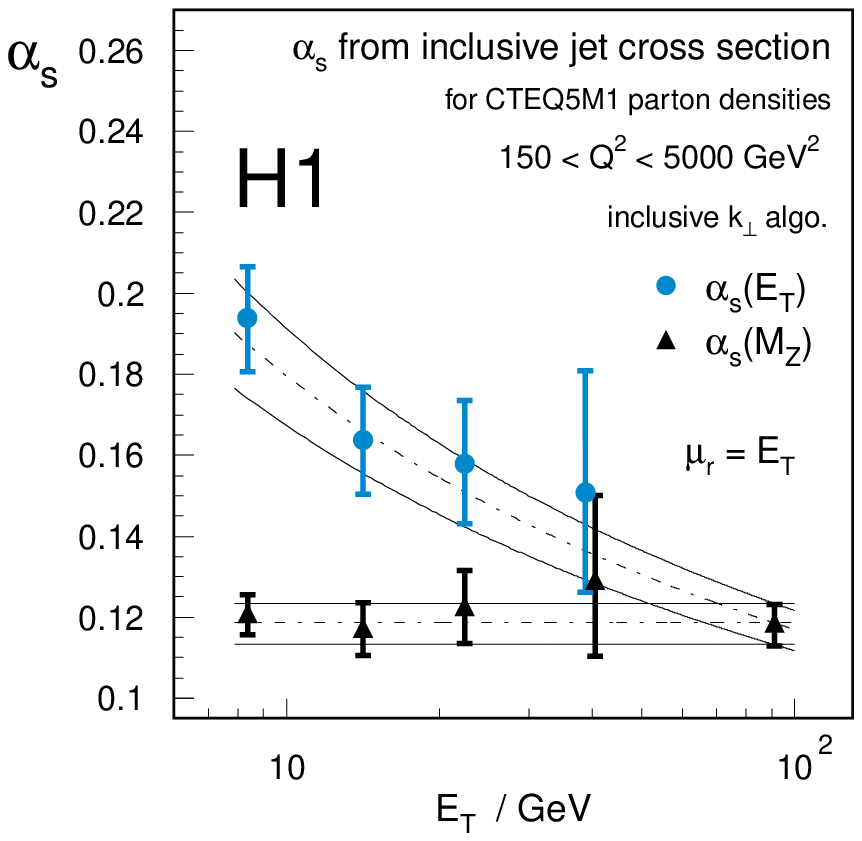,width=.8\textwidth}\hspace{.6cm}}
  \vspace{-.5cm}
  \caption{Determination of $\alpha_s$ from the inclusive jet cross section for
    the renormalization scale $\mu_r = E_T$. The results are shown for each
    $E_T$ value (circles) including experimental and theoretical
    uncertainties. The single values are extrapolated to the $Z^0$--mass
    (triangles). The final result $\alpha_s(M_Z)$ (rightmost triangle) is
    obtained in a combined fit. The curves represent the combined result
    and its uncertainties once at the $Z^0$--mass (lower curves) and once
    evolved according to the renormalization group equation (upper 
    curves).
    \label{fig=as_et} }
\end{minipage}
\begin{minipage}{0.44\textwidth}
  \raisebox{1.5cm}{ 
  \parbox{\textwidth}{ 
  \caption{The correlation of 
  $\alpha_s$ and the gluon density in the proton 
  from a fit to H1 jet and inclusive 
  DIS data for different values of the fractional momentum x of the 
  gluon.  The central fit result is 
  indicated by the full marker while the error ellipses show the 
  experimental and theoretical uncertainties.
  \label{fig=contour}}}} 
  \raisebox{-.2cm}{
\centerline{\epsfig{figure=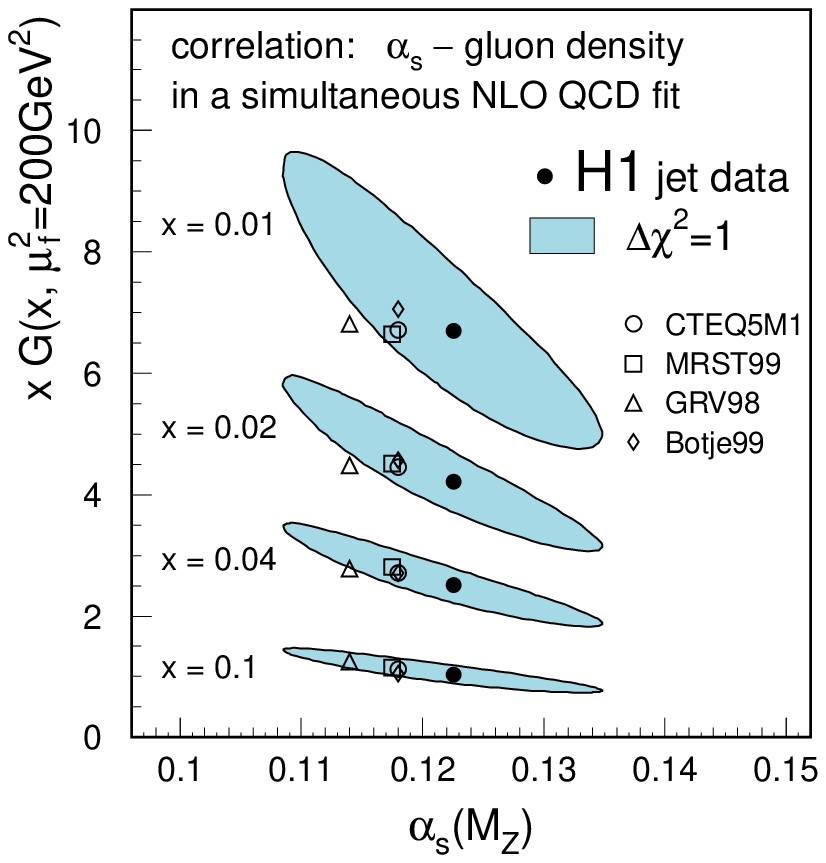,width=\textwidth}}}
\end{minipage}\hfill
\vspace{-.4cm}
\end{figure}

Jet production at large $Q^2$ and large $E_T$ has been studied by both 
the H1 and the ZEUS collaborations.  In this regime the choice of the 
renormalization scale ($\mu_{r}=\sqrt{Q^{2}}$, $E_{T}$) is of minor 
importance since scale dependencies of the cross sections are 
generally small if $\mu_r$ is large.  Hence NLO order 
calculations are expected to provide a good description of the data as 
can be seen from Figure~\ref{fig=incl_jets_kt} which shows the 
inclusive jet cross section $d^2\sigma/dE_TdQ^2$ measured by 
H1\cite{alphas}.  Jets have been defined using the inclusive 
$k_{\perp}$ algorithm\cite{incl_kt} in the Breit frame where the 
photon collides head-on with the incoming proton.  Over the whole 
analyzed phase space the DISENT NLO calculation\cite{DISENT}, using 
the CTEQ5M1 PDFs as input\cite{CTEQ5M1} and corrected for 
hadronization effects ($<10$\%), agrees perfectly well with the data.  
The size of the scale uncertainties together with the error 
contribution from hadronization corrections can be inferred from 
Figure~\ref{fig=incl_jets_kt} (right).  From these data, H1 extracts 
the strong coupling constant $\alpha_s$ in bins of $E_T$, chosen to be 
the renormalization scale.  The QCD predictions are fitted to the jet 
cross sections using the CTEQ5M1 parameterization of the PDFs and the 
strong coupling constant as the single free parameter.  The result is 
shown in Figure~\ref{fig=as_et} and clearly indicates the running of 
$\alpha_s$ according to the renormalization group equation.  A 
combined fit to all 16 data points shown in 
Figure~\ref{fig=incl_jets_kt} results in $$\alpha_s(M_Z) = 0.1186 \pm 
0.0030\; ({\rm exp.}) ^{+0.0039}_{-0.0045}\;
({\rm theo.})  ^{+0.0033}_{-0.0023}\; ({\rm pdf}) \;\;\;\;\;(\mu_r=E_T),$$
where the largest contributions to the experimental error come from 
uncertainties of the hadronic energy scale.  The theoretical error 
is dominated by the uncertainty in the hadronization corrections and 
the dependence on the choice of the renormalization scale.  The 
uncertainty in the knowledge of the parton density functions has been 
estimated\cite{alphas} using the correlated errors provided by a 
recent QCD analysis\cite{Botje} together with its global PDF-fits.  This error 
contribution is substantially reduced when using the dijet rate 
$R_{2+1}$ to extract $\alpha_s$ as done by the ZEUS 
collaboration\cite{zeus_r2}, leading to the preliminary result 
$$\alpha_s(M_Z) = 0.1166 \; ^{+0.0039}_{-0.0047}({\rm exp.}) 
^{+0.0055}_{-0.0042}\; ({\rm theo.}) ^{+0.0012}_{-0.0011}\; ({\rm 
pdf})
\;\;\;\;\;(\mu_r=\sqrt{Q^2}).$$

Both measurements of $\alpha_s$ depend on the knowledge of the parton 
content of the proton.  Vice versa, by using the best knowledge on 
$\alpha_s$ one can use \mbox{(multi-)}\-jet cross section measurements 
to extract the quark and gluon densities in the proton\cite{alphas}.  
A test of pQCD independent of data from other experiments, has 
been made by the H1 collaboration\cite{alphas} in performing a 
simultaneous determination of both quantities.  This is done in a fit 
to the inclusive jet and dijet cross sections together with the 
inclusive DIS cross section, where the latter is restricted to the 
kinematic range $150\le Q^2\le 1000$~GeV$^2$ and only constrains the 
quark densities in the proton.  The results of this fit are shown in 
Figure~\ref{fig=contour} as a correlation plot between $\alpha_s(M_Z)$ 
and the gluon density~$xg(x)$ evaluated at four different values of 
the gluon momentum fraction $x=$ 0.01, 0.02, 0.04 and 0.1.  While the 
present data do not allow a simultaneous determination of both 
para\-meters with competitive precision the sensitivity to the product 
$\alpha_s\cdot xg(x)$ can clearly be seen.

\section{Scale Ambiguities}

\begin{figure}[b]
\vspace{-.3cm}
\begin{minipage}{0.55\textwidth}
  \flushleft{\epsfig{figure=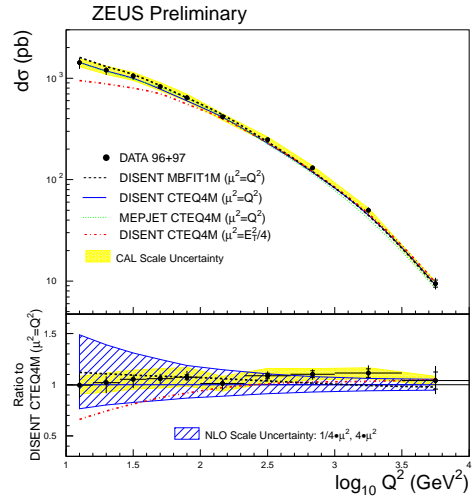,width=.95\textwidth}} 
  \end{minipage}\hfill \raisebox{0.0cm}{ \parbox{0.42\textwidth}{ 
  \caption{(a) The dijet cross section $d\sigma/d\log Q^2$ for jets in 
  the Breit frame using the inclusive $k_{\perp}$ algorithm.  The 
  points represent the data with statistical (inner bars) and total 
  (outer error bars) uncertainties.  The shaded band indicates the 
  systematic uncertainty due to the energy scale of the ZEUS 
  calorimeter.  While the theoretical predictions using
  $\mu_r=\sqrt{Q^2}$ describe the measured cross section rather well, 
  the dashed-dotted line representing a NLO calculation with
  $\mu_r=\sum E_{T,i}/2$, where $E_{T,i}$ are the 
  transverse jet energies of the jets, falls below the data for
  $Q^2<150$~GeV$^2$.  This can also be seen from (b) which shows the ratio 
  of the measured cross section to the NLO prediction using CTEQ4M 
  PDFs and $\mu_r=\sqrt{Q^2}$.}
\label{fig=zeus_dijet_xsec}}}
\end{figure}

The relevance of the choice of the renormalization scale at smaller 
$Q^{2}$ can be seen from Figure~\ref{fig=zeus_dijet_xsec}, showing the 
dijet cross section as a function of $\log Q^2$ as measured by the 
ZEUS collaboration\cite{zeus_dijet}.  As for the inclusive jet cross 
section, pQCD in NLO describes the data down to values of $Q^2\approx 
150$~GeV$^2$.  However, at $Q^2$ values below 150~GeV$^2$ scale 
uncertainties become large and the choice of the renormalization scale 
is no longer irrelevant: while NLO calculations using 
$\mu_{r}=\sqrt{Q^2}$ describe the dijet cross section  down to 
$Q^{2}\approx10$~GeV$^{2}$, this is no longer the case for $\mu_r = E_T$.

\begin{figure}[t]
\begin{minipage}{0.6\textwidth}
  \centerline{\epsfig{figure=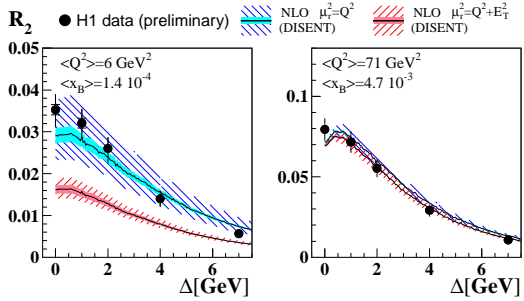,width=\textwidth}}
\end{minipage}\hfill
\raisebox{-0.2cm}{
\begin{minipage}{0.36\textwidth}
\parbox{\textwidth}{    
\caption{Dijet rate $R_2$ at two values of $x_{B}$, $Q^{2}$ 
  as a function of the transverse energy
  requirement $E_{T,1}>(5+\Delta)$~GeV on the jet with larger $E_T$; for
  the second jet the transverse energy $E_{T,2}$ must be above 5~GeV. 
  \label{fig=scale}}}
\end{minipage}}\hfill
\vspace{-.5cm}
\end{figure}

This issue has also been studied by the H1 collaboration\cite{roman} 
investigating dijet event rates, $R_2$, at low values of the 
Bjorken-$x$ variable $10^{-4}<x_{B}<10^{-2}$ and low $Q^2$, 
$5<Q^2<100$~GeV$^2$.  For different requirements on the transverse 
energies $E_{T,(1,2)}$ of the final state jets, NLO 
QCD calculations are confronted with the data using two different 
choices of the renormalization scale $\mu_r=\sqrt{Q^2}$ and 
$\mu_r=\sqrt{Q^2+E_T^2}$.  Figure~\ref{fig=scale} summarizes the 
preliminary result of this analysis in two representative 
($x_{B}$,$Q^2$)-bins.  At large $Q^2$ ($Q^{2}=71$~GeV$^{2}$) and large 
$x_{B}$ ($x_{B}=4.7\cdot10^{-3}$) the theoretical calculation is 
rather insensitive to the choice of the scale (Fig.~\ref{fig=scale}, 
right), and ---~in agreement with the results from the dijet cross 
section measurement at high $Q^{2}$~--- gives a good description of 
the data\footnote{This is true unless a symmetric transverse energy 
cut is applied where the NLO QCD calculation becomes infrared 
sensitive\cite{frixione}.  In this region resummed calculations are 
needed but not yet available.  Thus the presented measurement also 
provides an important reference for improved theoretical 
predictions.}.  In contrast, one obtains no safe theoretical 
prediction at low $x_{B}$, $Q^2$ (Fig.~\ref{fig=scale}, left): 
although data and NLO QCD calculation do agree for $\mu_r=\sqrt{Q^2}$, 
the large scale uncertainties lead to almost vanishing predictive 
power of the theory; choosing, however, a larger scale 
$\mu_r=\sqrt{Q^2+E_T^2}$ such that scale uncertainties become small, leads 
to clear disagreement between the measured event rate $R_2$ and the 
next-to-leading order QCD calculation, independent of the requirement 
on the transverse energy of the jets $E_{T,1}>(5+\Delta$)~GeV. This 
indicates the necessity to include higher-order contributions by 
introducing, for example, the concept of photon structure.  Choosing the 
same scale $\mu_r=\sqrt{Q^2+E_T^2}$, comparison of the data with the 
NLO calculations by JetViP\cite{JetViP} including a resolved photon 
contribution, indeed shows better, although not perfect 
agreement\cite{roman}.

\section{Probing Parton Dynamics}

\begin{figure}[t]
\begin{minipage}{0.49\textwidth}
  \centerline{\epsfig{figure=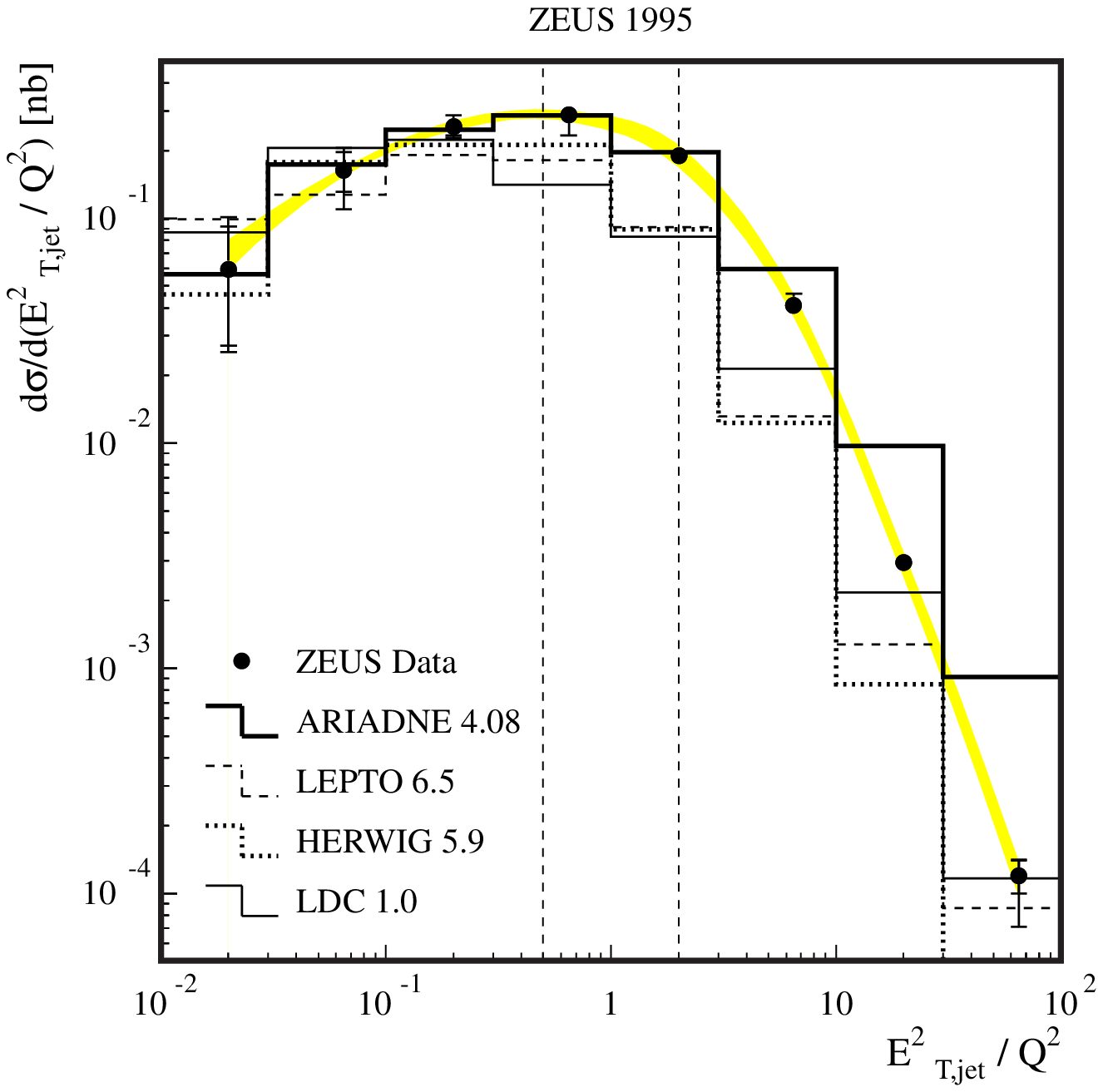,width=0.8\textwidth}}
\end{minipage}
\begin{minipage}{0.49\textwidth}
  \centerline{\epsfig{figure=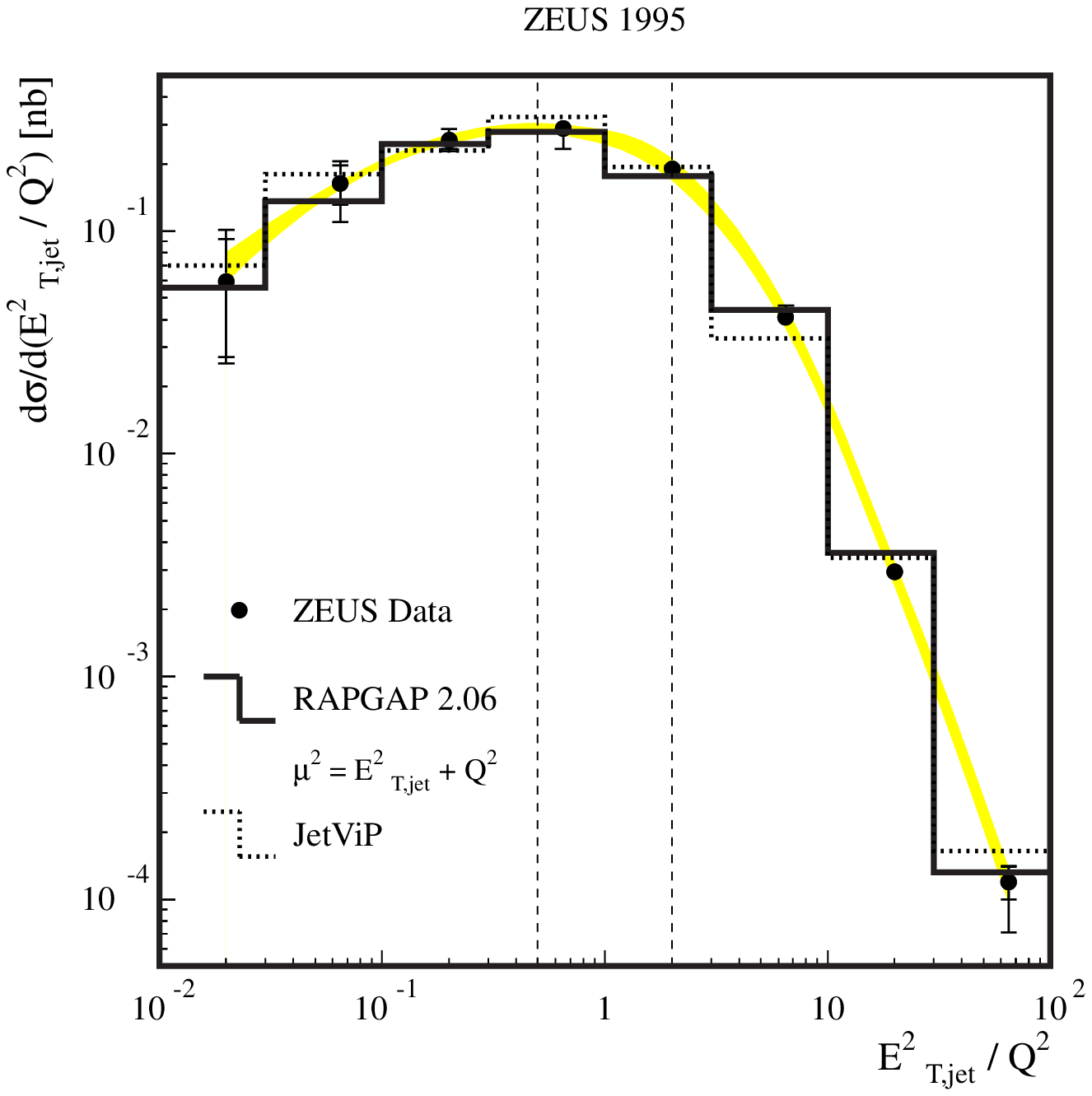,width=0.8\textwidth}}
\end{minipage}
\vspace{-.3cm}
 \caption{Forward-jet cross section as a function of $E_{T}^{2}/Q^{2}$ 
 in comparison (a) to several LO Monte Carlo predictions and (b) to 
 RAPGAP and a NLO JetViP calculation both including resolved photon
 contributions.  The shaded band corresponds to the energy scale 
 uncertainty of the ZEUS calorimeter.}
 \label{fig=zeus_fwdjets}
\vspace{-.8cm}
\end{figure}

In order to further investigate the interplay of the two scales 
$\sqrt{Q^{2}}$ and $E_{T}$, the ZEUS collaboration has 
studied\cite{zeus_fwdjets} the cross section for jets produced at 
large pseudo rapidities as a function of the ratio $E_{T}^{2}/Q^{2}$.  
When compared to LO Monte Carlo models (Figure~\ref{fig=zeus_fwdjets}) 
the data are well described for $E_{T}^{2}\ll Q^{2}$.  However, only 
those models which include non-$k_{t}$-ordered parton emission 
(ARIADNE\cite{ARIADNE}, RAPGAP\cite{RAPGAP}) reproduce the 
$E_T^2/Q^2\approx 1$ region; the full range can be described solely by 
models which include a resolved photon contribution as the RAPGAP 
Monte Carlo and a NLO calculation by JetViP. A reasonable description 
of these data is also found in a recent comparison\cite{jung} with 
Monte Carlo predictions based on the CCFM\cite{CCFM} evolution 
equation which, by means of angular-ordered parton emission, is 
equivalent to the BFKL\cite{BFKL} approach for $x\rightarrow 0$ while 
reproducing the DGLAP equations at large $x$.

\begin{figure}[t]
\begin{minipage}{0.57\textwidth}
\centerline{\epsfig{figure=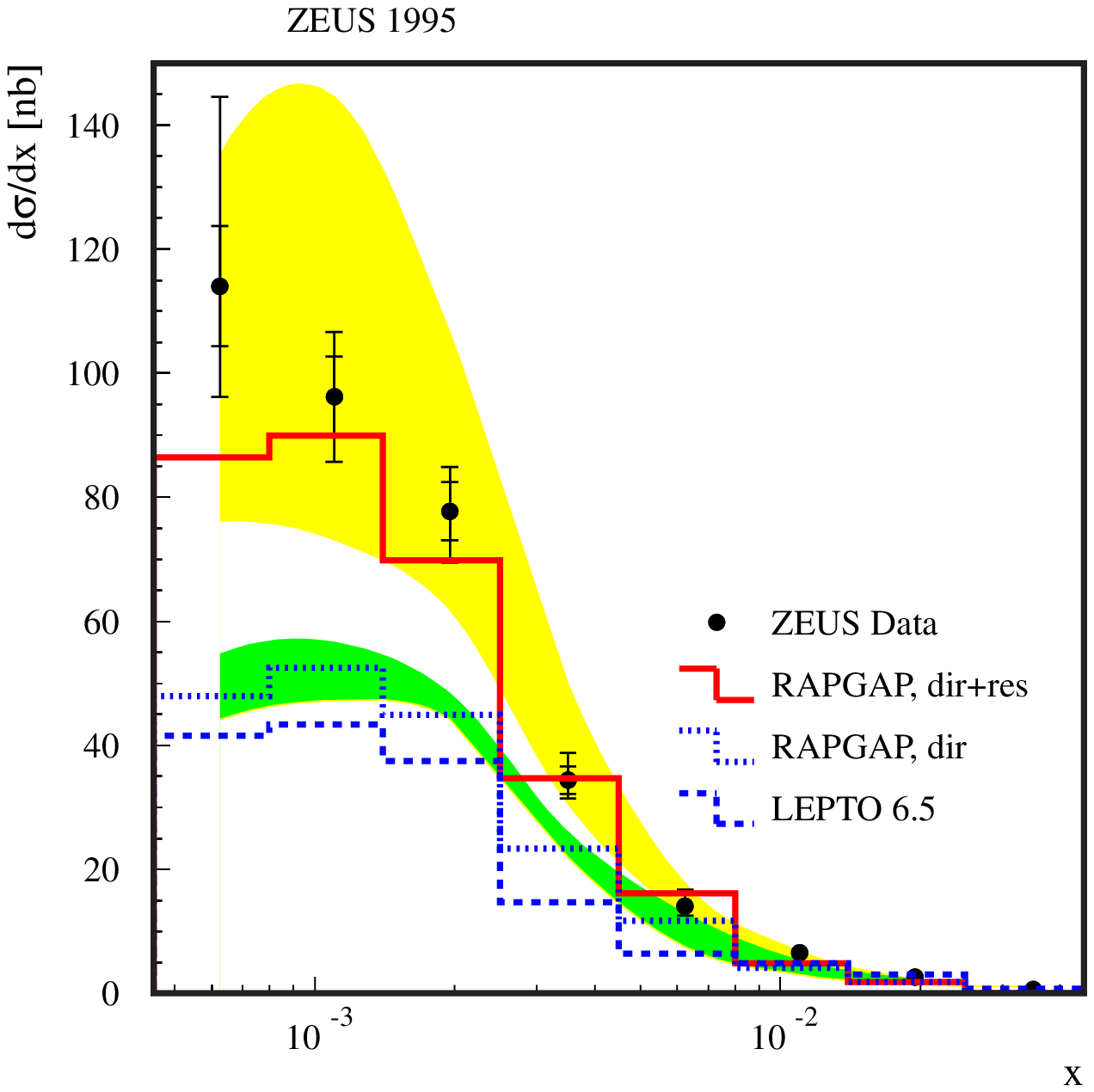,width=0.95\textwidth}\hspace{.3cm}}
\end{minipage}\hfill
\raisebox{0.10cm}{
\begin{minipage}{0.42\textwidth}
  \centerline{\epsfig{figure=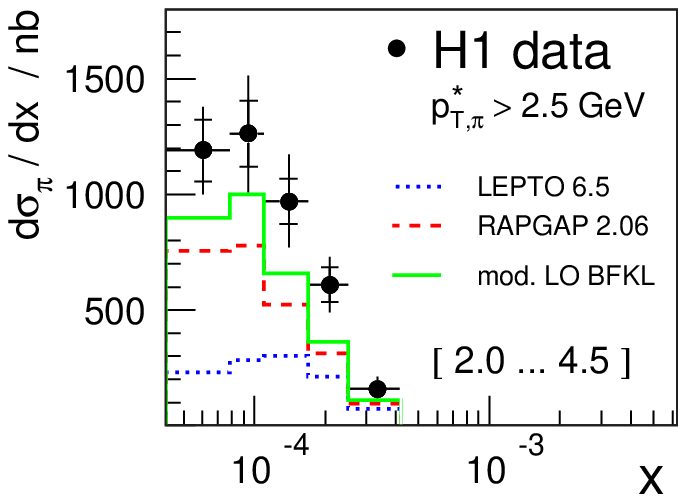,width=\textwidth}}
  \vspace{-.3cm} \parbox{\textwidth}{ \caption{$x_{B}$-dependence of 
  forward-jet (left) and forward-$\pi^{0}$ (right) cross section measured 
  for $E_T^2 \approx Q^2$.  The data are compared to Monte Carlo 
  models with and without resolved photon contribution as well as a LO 
  BFKL calculation (right figure only).\label{fig=zeus_fwdjets_x}}}
\end{minipage}}
\vspace{-.8cm}
\end{figure}

To explore a possible signature of BFKL dynamics, the $Q^{2}\approx 
E_{T}^{2}$ regime has been analyzed in more detail studying the 
$x_{B}$-dependence of forward particle production\footnote{Particles 
produced at very small polar angels w.r.t.\ the incoming proton 
direction.}.  Figure~\ref{fig=zeus_fwdjets_x} shows the 
$x_{B}$-dependence of the forward-jet and forward-$\pi^{0}$ cross 
sections in this region\cite{zeus_fwdjets,fwd_pi0} in comparison to 
Monte Carlo models with and without a resolved photon contribution and 
--- in case of the $\pi^{0}$ cross sections --- a LO BFKL calculation.  
The need for an additional contribution to the direct $\gamma 
p$-interaction can clearly be seen from these figures as the data are 
rather well described by models including a resolved photon component.  
Moreover, the $\pi^{0}$ measurement, which accesses very small polar 
angles, is actually best described by a BFKL-based leading-order 
calculation\cite{modBFKL} when taking hadronization corrections into 
account.  However, large theoretical scale uncertainties diminish the 
significance of these comparisons.

\section{Jets in Photoproduction}

\begin{figure}[t]
\begin{center}
\begin{minipage}{0.43\textwidth}
\centerline{\epsfig{figure=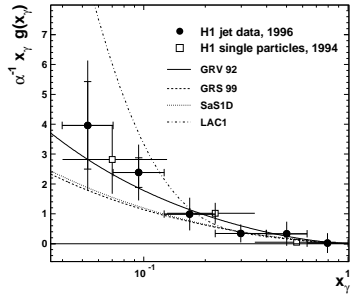,width=\textwidth}}
\end{minipage}
\raisebox{-1.2cm}{
\parbox{0.48\textwidth}{
\caption{The gluon distribution of the photon multiplied by 
$\alpha^{-1} x_{\gamma}$ as a function of $x_{\gamma}$ at a mean 
scale of $E_{T}^{2}=74$~GeV$^{2}$.}
\label{fig=gp_gluon}}}
\end{center}
\vspace{-0.5cm}
\end{figure}

The production of hard dijet events in photoproduction is dominated by 
resolved photon processes in which a parton in the photon with 
momentum fraction $x_{\gamma}$ is scattered from a parton in the 
proton.  Hence studies of dijet production can be used to investigate 
the hadronic structure of the photon at $Q^{2}\approx 0$.  In leading 
order the cross section is proportional to the photon flux 
$f_{\gamma/e}$ and maybe approximated by use of effective parton 
densities for the proton and photon such that $\sigma_{dijet} \sim 
f_{\gamma/{\rm e}}\cdot f_{{\rm eff},\gamma} f_{{\rm eff},{\rm 
p}}\cdot |{\cal M}|^{2}$, where ${\cal M}$ is the matrix element of 
the hard parton-parton scattering process and $f_{{\rm 
eff},[\gamma,{\rm p}]} = f_{{\rm q}/[\gamma,{\rm p}]} + f_{{\rm 
\bar{q}}/[\gamma,{\rm p}]} + 9/4\; f_{{\rm g}/[\gamma,{\rm p}]}$.  
With this relation H1 extracts the gluon density of the 
photon\cite{h1_g_gluon} in leading order pQCD using the knowledge of 
$f_{{\rm eff},{\rm p}}$ from DIS and $f_{[{\rm q},{\rm 
\bar{q}}]}/\gamma$ from the photon structure function $F_{2}^{\gamma}$ 
as measured in $e^{+}e^{-}$ collisions\cite{F2gamma}.  The result is 
shown in Figure~\ref{fig=gp_gluon} indicating a strong rise towards 
low $x_{\gamma}$.

\begin{figure}[t]
\centerline{
\epsfig{figure=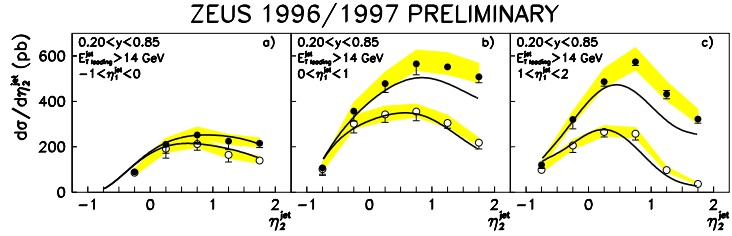,width=0.85\textwidth}}
\vspace{-.2cm}
\caption{The dijet cross section for $E_{T,jet}>14$~GeV as a function 
of $\eta^{\; jet}_{2}$ in bins of $\eta^{\; jet}_{1}$.  The solid data 
points correspond to the entire $x_{\gamma}^{obs}$ range while the 
measurement for $x_{\gamma}^{obs}>0.75$ is shown by the open points.  
The curves correspond to the NLO predictions and the shaded bands 
indicate the uncertainties related to the energy scale.}
\label{fig=zeus_gp}
\vspace{-.5cm}
\end{figure}

The determination of the gluon content of the photon is however 
strongly affected by non-perturbative effects and the treatment of 
background from soft underlying events\cite{h1_g_gluon}; it therefore 
can only be done using Monte Carlo programs which include 
phenomenological models to account for these effects but up to now are 
still restricted to leading order matrix elements.  Other dijet 
analyses from H1\cite{h1_virt_g_struct} and ZEUS\cite{zeus_gp_dijet} 
suppress these non-perturbative effects by harder $E_{T}$ requirements 
for the jets and thus investigate the parton distributions in the 
photon at high~$x_{\gamma}$~($x_{\gamma} \gtrsim 0.2$) where quark 
contributions dominate\footnote{Actually the dependence on 
$x_{\gamma}^{\rm obs} = [E_{T,1} e^{-\eta_{1}^{\rm \; jet}} + E_{T,2} 
e^{-\eta_{2}^{\rm \; jet}}]/2yE_{e}$ is studied, where $E_{e}$ is the 
electron beam energy.  In the leading order massless approximation 
$x_{\gamma}^{\rm obs}$ is equivalent to $x_{\gamma}$.}.  A comparison 
of the NLO prediction with the dijet cross section measured by 
ZEUS\cite{zeus_gp_dijet} for $E_{T,jet}>14$~GeV as a function of the 
pseudo rapidities of the two most energetic jets, $\eta_{1}^{\rm 
\;jet}$ and $\eta_{2}^{\rm \;jet}$, is given in Figure~\ref{fig=zeus_gp}.  
While the NLO calculation describes the data at large $x_{\gamma}\ge 
0.75$ rather well, a clear discrepancy is seen for the full 
$x_{\gamma}$-range if both jets are going in the forward direction 
(positive $\eta^{\rm \;jet}$).  This is where substantial 
contributions from resolved processes ($x_{\gamma}<0.75$) are expected 
suggesting that, in the kinematic region of the measurement presented, 
the available parameterizations of the parton density functions of the 
photon are insufficient\cite{zeus_gp_dijet}.

\section{Conclusion}

Measurements of $ep$ jet production rates have been presented in a 
wide range of the four-momentum transfer $Q^{2}$ and the transverse 
jet energy $E_{T}$ to study the dynamics of the underlying partonic 
interaction.  At large scales $\mu_{r}=\sqrt{Q^{2}}$, $E_{T}$ the data 
are successfully described by perturbative QCD allowing a 
determination of the strong coupling constant $\alpha_{s}$ and the 
gluon density in the proton.  Scale ambiguities have been studied and, 
as $Q^{2}$ gets smaller, sensitivity of the NLO predictions to the 
choice of the renormalization scale is observed.  At low $Q^{2}$ 
insight into the structure of the photon and effects of 
non-$k_{t}$-ordered parton evolution is revealed.

%

\end{document}